# Computing for Community-Based Economies: A Sociotechnical Ecosystem for Democratic, Egalitarian and Sustainable Futures

Robinson, K. P.[a]*, Eglash, R[a], Robert, L[a], Bennett, A.[b], Guzdial, M.[c] , Nayebare, Michael[a]

[a]*School of Information, University of Michigan, Ann Arbor, U.S.A.; [b]Penny W. Stamps School of Art & Design, University of Michigan, Ann Arbor, U.S.A.; [c]Electrical Engineering & Computer Science, University of Michigan, Ann Arbor, U.S.A.*

**Contact:** Kwame Porter Robinson; e-mail: kwamepr@umich.edu; mail: Attn: Kwame Porter Robinson, School of Information #4443, University of Michigan, 105 S State Street St., Ann Arbor, MI 48109-1285, United States

# Computing for Community-Based Economies: A Sociotechnical Ecosystem for Democratic, Egalitarian and Sustainable Futures

Automation and industrial mass production, particularly in sectors with low wages, have harmful consequences that contribute to widening wealth disparities, excessive pollution, and worsened working conditions. Coupled with a mass consumption society, there is a risk of detrimental social outcomes and threats to democracy, such as misinformation and political polarization. But AI, robotics and other emerging technologies could also provide a transition to community-based economies, in which more democratic, egalitarian, and sustainable value circulations can be established. Based on both a review of case studies, and our own experiments in Detroit, we derive three core principles for the use of computing in community-based economies. The prefigurative principle requires that the development process itself incorporates equity goals, rather than viewing equity as something to be achieved in the future. The generative principle requires the prevention of value extraction, and its replacement by circulations in which value is returned back to the aspects of labor, nature, and society by which it is generated. And third, the solidarity principle requires that deployments at all scales and across all domains support both individual freedoms and opportunities for mutual aid. Thus we propose the use of computational technologies to develop a specifically generative form of community-based economy: one that is egalitarian regarding race, class and gender; sustainable both environmentally and socially; and democratic in the deep sense of putting people in control of their own lives and livelihoods.

## I. Introduction

**A. Overview**

As Graeber (2001) notes, there are two meanings to the word "value". One is the normative or ethical sense of the word, as in "our most cherished values". The other is in the economic sense we use it here. Graeber suggests that the ways in which neoliberal economic theory assumes money to be the universal representative of value blinds us to important distinctions in value forms.

In contrast, we utilize the framework of generative justice (Eglash 2016) to suggest where and how value forms arise. Eglash (2016) defines "generative justice" in terms of three categories: ecological value, labor value, and social value. In all three cases, sustainable systems return, or *circulate*, value back to their sources and crises are created when value is expropriated with little return to them. Value circulation is an important concept in many natural science fields; scholars such as Schrodinger (2012) have pointed to the profound nature of regenerative cycles in understanding biological life. There are different names for it at different scales — autocatalysis for biomolecules, self-reproduction for organisms, self-assembly for ecosystems — but they are all referring to the loop from self back to self. The multiple scales enable "recursive depth": there is a nesting of cycles within cycles. These deeply nested, circular flows of matter, energy and information are the basis for the self-sustaining, flourishing, adaptive structures we call life (Cornish-Bowden and Cárdenas, 2020).

Generative justice applies a value distinction across 3 domains: ecological, labor, and social. The phrase "unalienated value" is recursively defined as that which makes possible just and sustainable value generation. In the ecological domain unalienated value might manifest in organic agriculture using generative recycling. In the labor domain, Ostrom (1990), Cox et al. (2010), Bedford and Cheney (2013) and others have shown that Indigenous traditions created the means of cycling labor value back to the people (and ecosystems) that sustained them, all examples of unalienated value circulation. In the social domain, value circulation includes cultural meaning, creativity, conviviality, solidarity, trust, and commitments to democratic processes (Pazaitis et al., 2022). The potential damage, *alienated value*, often carried out by social media is essentially the extraction of social value from its proper cycles. Using AI, data surveillance, recommendation algorithms, disinformation, polarization, and other means, the human sociality that normally enriches our own communities is extracted for purposes of wealth accumulation and manipulation of voting, banking, consumption and other behaviors (Wagner and Eidenmuller 2019; Pellegrino et al. 2022; Zuboff 2015).

What differentiates Generative Justice from other theories is that egalitarian interaction should proceed bottom-up and across social, ecological, and labor domains. Firstly, the participants organize themselves across the full cycle of value generation and circulation; and secondly, the value remains within the participant community and while it can be shared the value is not expropriated away. In the unalienated form, cyclic generation, sharing in self-generated value, is a kind of *unalienated* value whereas extracting self-generated value by not sharing in its creation — the alienated form of value — can break egalitarian cycles.

We posit that a community-based economy, given the appropriate technical and political resources, could replace value extraction with circulation. By developing computational technologies for regenerative cycles, these economies would reduce environmental damage, reduce wealth inequality, and establish more democratic ways of life.

To illustrate how these principles of generative justice can be applied, figures 1 and 2 show a comparison between value flows in two cases. Figure 1 is value flow chart for traditional Iroquois economy, adapted from Eglash (2016). At the smallest scale, the traditional "3 sisters"--beans, corn and squash--create a regenerative cycle among non-human agents. Bean root nodules contain nitrogen-fixing bacteria that replenish the soil; corn provides climbing structure for beans; squash has spiny leaves that prevent weeds, pests, and loss of soil moisture (Mt. Pleasant, 2016). That cycle is nested within a labor value cycle: work collectives ensure each household farm has sufficient workforce, with harvests divided between a shared commons and individual household storage (Abler 1993). That cycle is nested within an intra-tribal political structure balancing power between clan mothers and the tribal council (thus establishing women's voting rights centuries before any European nation; and inspiring US suffragettes in their voting rights movement (Geredien, 2018). And that cycle is nested within the inter-tribal Iroquois confederacy; a structure that US founding fathers cited as inspiration in their creation of the constitution (Miller, 2010). It provides an impressive model showing that nested cycles of value regeneration are possible, not only creating an economy without unjust extraction or alienation, but inspiring the birth of democracy.

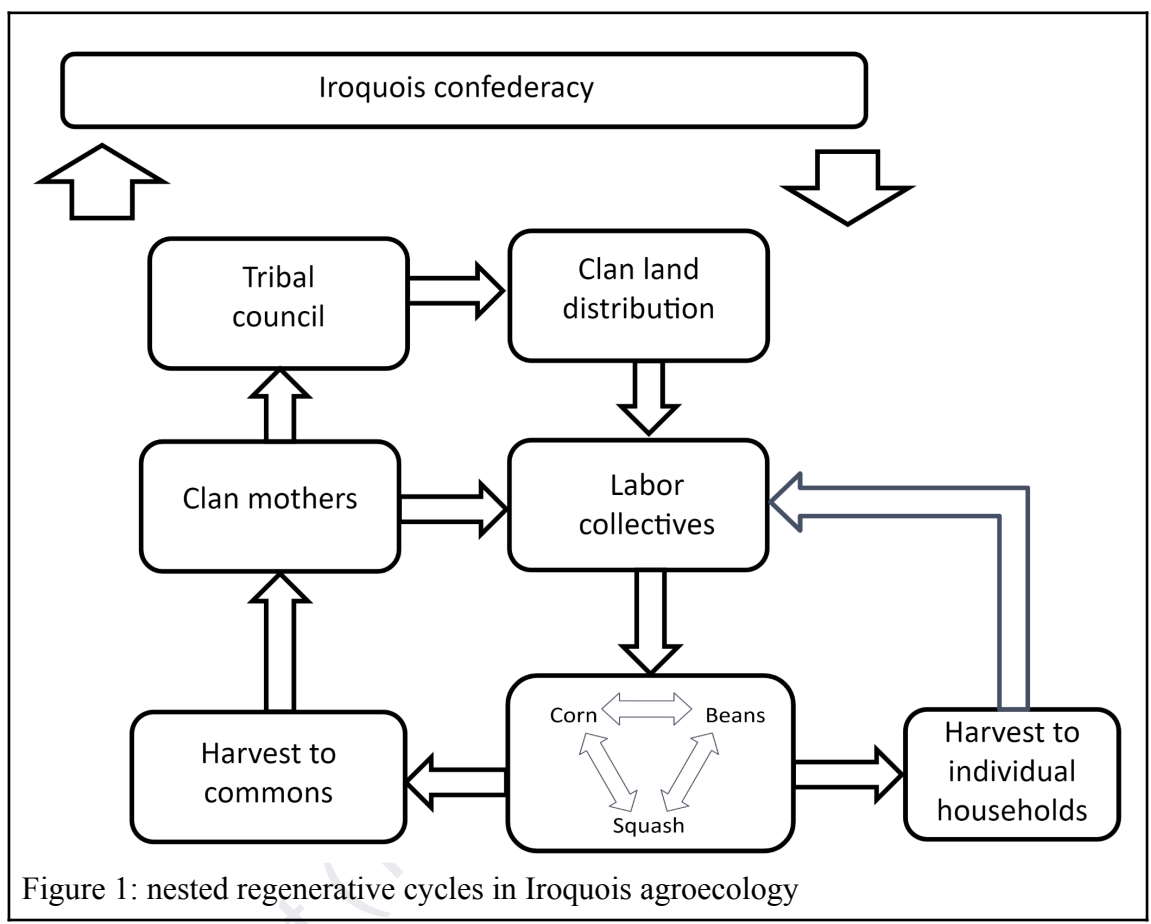

Figure 1: nested regenerative cycles in Iroquois agroecology

In contrast, the Arduino production network in figure 2 has a combination of regenerative cycles and extractive value flows. Arduino is an open source, user-friendly microprocessor board, widely celebrated for its pivotal role in democratizing lay access to physical innovation (Kushner, 2011). At left in figure 2, electronic component manufacturing, especially for silicon chips, utilizes mass production systems often associated with pollution, labor exploitation and social distress (Smith et al., 2006). We have drawn the arrows for alienated value flow as solid single lines to indicate their reduced state.

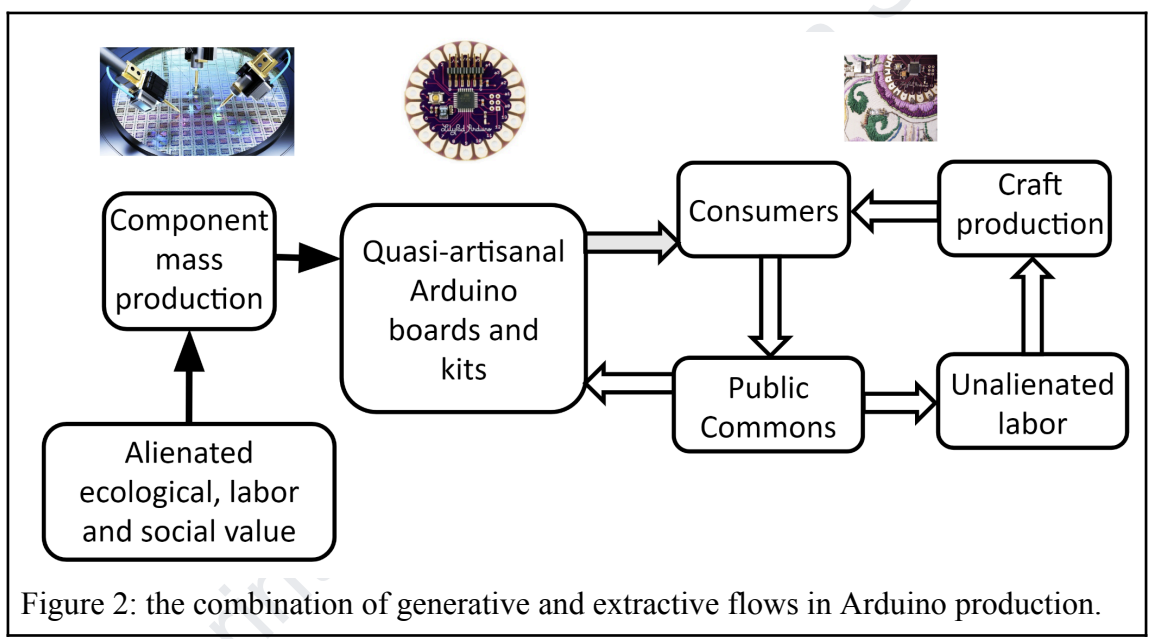

Figure 2: the combination of generative and extractive flows in Arduino production.

At center, mid-sized companies such as Arduino in Italy, SeeedStudio in China, SparkFun in Colorado, and Adafruit in Brooklyn have created a "quasi-artisanal" mix of hand-crafting and assembly line production for unique Arduino boards and kits. It allows rapid innovation and response to user requests that fits their niche in the "maker" market of DIY hobbyists. Eschewing the traditional focus on proprietary legal protections and profit-first decisions, they have flourished by developing a combination of open source access, socially progressive business policies, and deep ties to the "maker" movement communities in which they are beloved (Eglash 2016).

The gray arrow at center in figure 2 indicates a quasi-unalienated status: these hobbyist-serving companies are not worker-owned, but otherwise do an admirable job in resisting many of the pressures toward value alienation (Buechley, 2009; Eglash, 2016; Lindtner et al., 2016; Lindtner and Lin, 2017; Rimmer, 2021). That is at least in part because of the unique character of their customers: in particular the use of Arduinos and related kits for DIY pollution detection, anti-surveillance devices, subversion of corporate anti-repair and anti-hacking designs, and other counter-hegemonic applications. Hence the open arrow returning value to the mid-sized companies at center: new ideas for kits come from these consumers (Hamalainen and Karjalainen, 2017). The cycle of four open arrows at the right in figure 2 indicates the regenerative nature of the commons-based sharing of open source code and hardware designs within the Arduino user community. Indeed the broader open source movement has spawned a wide range of similar peer to peer networks, cooperatives and artisanal collectives (Benkler 2007; Bauwens and Pazaitis 2019; Garvin et al. 2023).

The contrast between figures 1 and 2 offer three insights. First, they illustrate how the abstract principles and terminology of generative justice can be applied to concrete cases. An entire economy in which the regenerative flows have "recursive depth", requiring cycles within cycles, is illustrated by the Iroquois example in figure 1. Second, figure 2 is a cautionary reminder that in the contemporary context, even purely regenerative cycles (such as the peer-to-peer network of the Arduino open source commons) will almost certainly have ties somewhere in its supply chain or lower-level infrastructure to processes of value alienation (such as the chip industry at the left in figure 2). Third, on a more optimistic note, even those compromised regenerative cycles can have a positive impact, supporting the existence of hybrid, quasi-unalienated forms, such as figure 2 center.

This paper asks how impactful these regenerative cycles can be. Could multiple regenerative cycles, deliberately structured by democratic process, extend across the depth and breadth of the entire community, as we saw in the Iroquois case, but applied in the contemporary context? We will review some of the evidence that computational supports, allowing these three forms of unalienated value — intersecting in myriad ways — can embed generative justice in localized economies (Krueger et al. 2023a,b; Dillahunt et al. 2023; Eglash et al. 2024). We will refer to such systems as *community-based economies* or CBEs and examine their potential to create more sustainable, equitable, and democratic economies.

### B. Assessing community-based economies as generative justice

The basis for assessing community well-being encompasses a wide range of parameters, including health, housing, education, income, worklife, environment, civil rights, and so on. For example in 2015 the United Nations adopted 17 sustainable development goals (Biermann et al. 2022). A community-based economy (CBE) shares these goals, but frames a particular solution for achieving them: one in which the ownership of wealth generation is rooted among the local community members, rather than centralized in large corporations or state ownership (Newman 1986). CBEs also emphasize localization of consumption, and other means of developing horizontal ecosystems in which sharing, recycling, upcycling, and communal activities can provide cooperative alternatives to the competitive individualism, hyperconsumption, planned obsolescence, fast fashion and other trends driving much of the environmental damage and wealth inequality that threatens community well-being (Acquier et al. 2017).

As noted above, our theoretical foundation for these features are derived from the framework of generative justice: "The universal right to generate unalienated value and directly participate in its benefits; the rights of value generators to create their own conditions of production; and the rights of communities of value generation to nurture self-sustaining paths for its circulation" (Eglash 2016). The phrase "unalienated value" is recursively defined as that which makes possible just and sustainable value generation. It is thus analogous to the circular flows of autopoiesis that underlie all organic life (Cornish-Bowden and Cárdenas, 2020), but here applied to sociotechnical production (e.g. Krueger et al., 2023a,b; Dillahunt et al. 2023; Eglash et al. 2024).

The opposite of generative is extractive, and the same three categories of ecological value, labor value and social value are the primary locations in which value is alienated from what would normally be regenerative cycles. Ecological value extraction includes clear-cutting forests, overfishing oceans, toxic disposal and other mass production forms that colonize the flows that normally allow nature to regenerate (Austin et al. 2017). In similar ways, social value extraction is carried out when online social media allows the colonization of our social networks, including data surveillance, consumption rerouting, disinformation dissemination, political polarization, and other ways in which the human connectivity that normally enriches our own communities is extracted for wealth accumulation by corporations (Pellegrino et al. 2022; Zuboff 2015).

Generative cycles for unalienated value in the social domain, may seem to blur the distinction between this economic definition of "value" and its ethical meaning ("our cherished values"). Graeber (2001) points to this ambiguity as central to contemporary destructive processes. Marketing science has advanced to the point where the worst polluters on the planet can convince the public of their love of nature; labor exploitation can appear to "treat workers like our family" and so on. No matter what rhetoric disguises it, when value is extracted in unsustainable ways, this action is breaking the generative cycle, causing degraded environments in ecology, exploitation in labor, and domination in society. Conversely, a successful community-based economy might (hopefully) be claimed a success by many ideologies: political conservatives who see small family-owned business as embodying "family values", political radicals who embrace labor autonomy, and all those in-between.

**C. Bottom-up vs top-down**

While various ideologies may claim success in establishing values, the key lies in how systems preserve and enhance generative cycles. Here the contrast between preserving generative cycles and *historical labor extraction* is instructive. The consequences of labor value extraction were defined by Marx (1844) using four dimensions: alienation of the worker from products, from work practices, from our sense of creative agency, and from other workers. As noted above, a top-down approach that continues extracting value, but passes it on to a centralized state for redistribution, has in many cases (USSR, China, etc.) failed miserably (Eberstadt, 2017; Mazurski, 1991). That is to say, achieving a generative economy is just as challenging under the extremes of forced collectivism as it is under the extremes of free market competition. Because unalienated value is relational--it can only be created and circulated through voluntary common-pool sharing, reciprocity, fair trade, and other solidarity mechanisms--it requires "bottom-up" organization; a balance between the free agency of individuals (in contrast to forced collectivism), and opportunities for mutual aid (in contrast to free market competition). Community-based economies are one of the most promising means for achieving this goal, because their localized structure can facilitate the return of ecological, labor, and social value flows in the most direct ways, to those mostly directly affected. The smaller size[1] of community-based economies better facilitates bottom-up democratic processes. Of course those will work best if there are similar circular flows at regional, national and international scales, but that is beyond the scope of this paper.

Challenges to transitioning to generative forms of community-based economies in contemporary contexts include deeply entrenched structures of segregation and stratification, highly privatized infrastructures and resources, damaged ecosystems, and significant cultural conflicts. In this paper we consider the potential role of computational technologies to overcome these challenges, and develop generative (as opposed to extractive) community-based economies. We describe this as both the desirable end-state, as well as a practical means of transition that upholds these values at every step. In other words, we propose transitions to sustainable, economically egalitarian and deeply democratic communities through applications of computational technologies that are designed for minimizing value extraction, and maximizing the return of unalienated value to the ecosystems, labor systems and social systems in which they were created.

---

[1] Efforts such as MIT's FabCities initiative (described below) typically focus on large urban cities such as Barcelona as the definition of size. They describe larger scales in different terms (e.g. regional "hubs" between community-based economies).

The problem of value extraction leading to unsustainable degradation can be solved by two approaches. The top-down approach, famously described by Hobbes' *Leviathan*, requires a centralized authority to prevent unsustainable extraction and conflict, or in the case of Marx, to redistribute the extracted wealth. Marx's 1848 Manifesto of the Communist Party calls for strict centralization of "all instruments of production" (factories, machines, agricultural estates, mines, etc.) as well as finances, communication, transportation, and even the workforce--an "industrial army" --in the hands of the state (1974, pp. 86-87). Selucky (1974) examines the deeper rationale behind this command economy: unless "autonomy of production units is abolished" skilled workers would never agree to the same pay as unskilled, managers would refuse to hire the poor, companies would be in competition with each other, and capitalism would simply rebuild itself after the revolution.

But simply proclaiming that an economy will be organized bottom-up, and nurture autonomy rather than abolish it, does not make any of these challenges vanish; to the contrary they take center stage. Even for the small scale, cohesive traditional societies in the commons-governance framework examined by Ostrom and others, this is a non-trivial task. Ostrom delineates 8 principles required, such as monitoring to ensure adherence to rules, democratic control over rule modification, and so on. The goal of contemporary community-based economies, in which we are scaling up from a fishing village to a modern city, and thus striving to achieve unalienated value flow across multiple domains from labor to transportation to health, requires a far more complex set of mechanisms.

The trade-off--sacrificing the simplicity of a top-down command economy for the emergent complexity of bottom-up approaches--is worthwhile if it results in an economy that is democratic, egalitarian, and sustainable. In this paper we examine the ways in which computation can help to resolve this complexity challenge. That includes the variety of forms that worker autonomy and ownership of production might encompass at bottom, the mid-range civic institutions such as mutual-aid financing, and state support for required legal structures. Thus "bottom-up" is more useful as a conception of directional flow--the recursive circulation of unalienated value, from its sources of generation at the grassroots and back to them, minimizing extraction whenever possible--than it is creating a reductive binary distinction from top-down.

**D. Terminology for Related Approaches to Community-based Economies**

 The terminology above--geographically bounded community, value generation with minimized extraction, and bottom-up value circulation--are fundamental concepts for describing what we mean by "community-based economy". There are a wide variety of closely allied terms, each with advantages and gaps. "Commons-based peer production" is not bounded geographically; it includes processes that might be global in scope. But it best represents the aspects of open source production that are needed for community-based economies (Benkler 2007; Bauwens and Pazaitis 2019).

"Cosmolocal" is broader still, so much so that at times it is difficult to distinguish from ordinary capitalism (see Li 2021 or Kostakis and Tsiouris 2024). But it highlights concrete actions such as the maker movement that might be missed in more restricted formulations. "Circular economy" or "regenerative economy" is often focused solely on environmental sustainability, putting concern for wealth inequality and democratic control as a secondary concern at best (Schröder et al., 2020). "Solar punk" is a relatively new term stressing the DIT ("do it together") maker ethos, and technology appropriation for oppositional politics (Eglash et al., 1994), rather than a predefined systemic vision (Reina-Rozo, 2021). Finally, "solidarity economy" is nearest to our set of definitions. In most cases (e.g. Miller 2010, Giovannini, 2020) it refers specifically to networks of worker-owned enterprise, cooperatives, peer to peer production, collective governance of resource commons, and civic institutions for collaboration and mutual aid. The phrase "community-based economy" itself has been used in widely varying ways. Gibson-Graham (1996), a joint pseudonym for Julie Graham and Katherine Gibson, established the branch of that literature most closely allied to our use here, and their research collective at communityeconomies.org continues to publish in that framework.

## II. Five Strategies for comparing community-based economies: comparing before and after computing

### A. Community-based economies prior to computing

Prior to computing, attempts to shift economies to more localized and egalitarian systems developed five fundamental strategies that are still in use today: decentralized worker collectives, centralized worker-ownership, socio-legal supports, alternative exchange instruments, and democratic control of common-pool resources.

*Decentralized worker collectives*

While capitalism stresses the individual, and communism stresses the collective, "solidarity" can be defined as mutual support between individual agency and collective action. The phrase "solidarity economy" was first used as a description for community-based economies in 1937, when Felipe Alaiz advocated for a federation of autonomous municipalities that could support exchanges between worker collectives in urban and rural areas during the Spanish Civil War (Cleminson 2012). Since then, worker collectives have remained an important part of the broad range of strategies that the term "solidarity economy" now encompasses. The advantages of worker collectives can be summarized in terms of Marx's four dimensions: the ability to collectively determine the products; to use preferred styles of production; to add meaning and agency to the work process, and to work collectively with others in a spirit of mutual aid. But without the support of a broader solidarity economy, they may fall short of these goals (Rothschild-Whitt and Whitt 1986).

*Centralized worker collectives*

Peredo and Chrisman (2006) define community-based enterprise as "a community acting corporately as both entrepreneur and enterprise in pursuit of the common good". They use the Mondragon federation in Spain as a contemporary example: although it began as a small worker collective, by pooling profits for allowing new ventures, it became a community-wide economic collaboration of over 80,000 workers, fostering improvements in social capital such as education, environment and democratic governance.

*Socio-political supports*

Piore and Sabel (1984) described networks of small-scale manufacturing and artisanal production in northern Italy. These Flexible Manufacturing Networks (FMN) were extolled for their high wages, worker education and other social returns, but development attempts to replicate them elsewhere have largely failed. Zabin and Ringer (1997) suggest two factors. The first is social: that this region had long-standing networks of family ties that facilitated trust in collaborations. The second is legal: local laws prevented the kinds of low-paid gig work that flexibility has engended elsewhere, and made investments in reskilling workers the only option. A recent example of broad based socio-political support can be found in the concept of *partner state* (Bauwens et al 2019b), where the state facilitates commons creation with citizens and their communities through peer to peer economies and partnership, as lived in, for example, the Emilia-Romagna region of Italy (Bardia 2023). Other examples of socio-political dominant approaches and theory may be found in Kostakis and Bauwens 2014, or even direct democracy assemblies as advocated by Bookchin and Castordias (Tarinski 2016). Political commons governance (c.f. Ostrom's institutional analysis and development framework) is another emerging area (Papadimitropoulos 2020b). Below we examine whether computing technologies can fulfill these objectives in areas where such socio-legal support is absent or diminished.

*Alternative exchange systems*

These include localized currency, in which a community essentially prints their own form of money; time banks, in which hours of work are exchanged; barter systems; and gift economies. Because these have largely shifted to online exchange systems, we detail them in the next section.

*Democratic process for common-pool resources*

Modern solidarity economy movements have critiqued representative governments' response to local needs, and developed direct democracy approaches to shared public resources. Some are closer to Ostrom's classic studies of community-based natural resource management (Smith 2019). Others include participatory budgeting, where lay citizens are invited to make decisions on city or regional development spending (Touchton et al. 2023). Specific designs are required to overcome inclusion barriers for gender, class, and cultural differences (Wampler and Touchton 2019).

*Summary*

In sum: all means for developing community-based economies can be understood as methods for preventing value extraction, and ensuring that those generating the value--whether non-humans in nature, workers performing labor, or groups of people interacting in social and civic spheres--are able to control their own contexts of production, and thus create value without alienation. This allows them to circulate that unalienated value back in a regenerative cycle that enriches those forms of production. Because alienation is defined as relational--recall Marx's four dimensions--the term "unalienated" in this context is understood to mean sustainable, egalitarian, and democratic practices.

### B. Computing within Community-based Economies

While technologies such as the printing press, accounting machines and the like have been instrumental to community-based economies, the introduction of silicon chips, online networks and machine learning represent a vast shift, both in terms of what threatens community well-being as well as its potential solutions. However we can still use the same five categories to examine computational means for developing community based economies.

*Decentralized worker collectives and Peer Production*

Computation is a fundamental aspect of contemporary peer production and decentralized work. It enables long-distance communication, provides templates for work arrangements, and structures platforms for organizing labor and managing relationships (Benkler and Nissenbaum 2006). Examples include GitHub for software development (Tsay, Dabbish, and Herbsleb 2014), decentralized hardware blueprints (Open Source Ecology 2023) for self replicating machine ecosystems, or Rhys Jones' RepRap project  (Jones 2011). In social and well being, information sharing across health forums enhances access to health resources, transferring knowledge from urban to rural users (Gho 2016) through peer production of social capital. Many other examples exist. Benkler's *Wealth of Networks* (2007) is a foundational text for understanding these peer to peer interactions as a grassroots alternative to corporate domination..

However, the ease and speed at which technology allows these platforms to be established often leads to a lack of nuanced governance and oversight of social relations, resulting in norm violations and failures (O'Neil 2013). The use of computation at a wider social distance, such as weak ties, can exacerbate social problems, underscoring the importance of community guidelines for establishing stronger relationship ties. This raises questions about how expectations change depending on distance and relationship of peers, and of the limitations of pre-established community norms or templatized work arrangements that do not address these nuances. For example, the trajectory of peer produced artifacts are highly sensitive to expectations of community (Arazy 2020), despite codes of conduct (that are often violated). Additionally, the co-optation of peer production by surrounding capitalism further exacerbates norm violations and failures (Birkinbine 2020). The copy and pasting of incomplete production structures and implicit social assumptions, facilitated by computation, perpetuates many of these issues. Frameworks to address these challenges by stressing deliberative relationships and localized modes of production include prefigurative approaches (Fians 2022), peer to peer processes (Kostakis, Vragoteris, and Acharja 2021; Rigi 2013), and cosmolocalism (Ramos 2021). Our argument for Computing for Community Economies, presented in Section 3, takes a specific prefigurative approach embedded in the participatory design of automation technology and artificial intelligence applications, while providing an operationalization thereof.

*Centralized worker ownership: platform cooperatives*

In the industrial age, Mondragon in Spain offered one of the best examples of centralized worker-ownership in the context of community-based economies. In the post-industrial age, the computational equivalent would be platform cooperatives. In contrast to decentralized networks, where organizations may be small and more easily self-determined, collective ownership such as platform cooperatives must combine democratic processes with the organizational demands for large, complex operations. Computation has many advantages for this application, and in particular automation technologies are often used to aggregate, classify, and track data, with multiple quantitative optimization goals of reducing costs, increasing efficiency and facilitating transaction throughput (Lampinen 2018).

Within HCI literature, platform cooperatives fall under the broader category of civic technology, which encompasses computational forms that trouble corporate domination and support community-based alternatives (Scholz and Schneider 2017). In one of the broadest reviews of cooperatives to date, Lampinen et al describe computing artifact ecologies, cobbled together assemblages of email and other ICT tools, as periodically negotiated, situational, and discarded, through the course of periodic group reflection (Lampinen 2018). Rose (2021) distinguishes between two types of "non-corporate platforms": the platform cooperative that offers products or services for payment (but enjoys worker ownership), and FLOSS platforms that fulfill non-business functions (such as direct democracy) and operate more like an open source project than a business. Many cooperatives wish for greater inter-cooperation in "today's fragmented world of work" (Lampinen 2018). Computing for community economies uses the principles of generative justice to simultaneously synthesize these concerns using a pre-figurative approach and represents one approach for community computation using participatory design to construct and inter-link social, environmental, and economic circuits.

*Socio-political supports*

We define the cultural, political, legal, and institutional frameworks that enable and promote solidarity economies in a given region as socio-political supports. Many of these are historically embedded. For example, the Basque region of Spain, the site of a rich anarchist tradition (Ealham 2010), has been home to several notable examples: the origins of several solidarity economies in the 1930s (Ealham 2004, Bookchin 1994) as a counter revolutionary reaction to Spanish republic monarchy, the collective Mondragon movement in the 1950s (Bateman 2015; pg 5) utilizes legal incorporation to receive state support, and Barcelona's  solidarity organizations in late 2010s (Giovannini 2020).

Today Barcelona is often cited as the leading case for computing support for solidarity economies, including platforms for cooperative marketing, fabrication, democratic decision making and other functions (Wouter 2022).

One does not have to have a deep history to introduce contemporary socio-political supports. Sometimes viewed as a set of tactics[2], socio-political support has largely been defined by four approaches: 1) State supported economic protections such as exemptions from monopoly laws, facilitation for participation in conventional economies, and other policy changes (Hanisch 2013, Meister 2020, Crivellaro 2019, Manuel 2020, Lémbarri 1995) . 2) digital support for social services such as volunteer organizations, immigrant settlement organizations, public transportation, and other exemplars (e.g Vlachokyriakos 2018, Crivellaro et al. 2019 of Open Labs in the UK; drawing on conceptions of "infrastructuring" from Star 1999); 3) networking to support public access to fabrication knowledge sharing and physical making resources, such as the FabCity concept (Rumpala 2023). 4) Building public support for local creative capital, sense of place, and other less tangible forms (Jones 2021). Among these cases, the action needed at the level of the state is sometimes to add new support for existing efforts, other times to remove legal and policy blocks that prevent the formation of solidarity efforts.

*Alternative exchange systems*

Alternative exchange systems are predicted on reciprocal trust that can be difficult to establish and sustain. Where these systems have survived tend to contain one or more attributes of strong pre-existing relationships of trust (Satori 2020; c.f. Sardex), low fiat money liquidity (Ruddick et al. 2015, Mattsson et al. 2022, Zeller 2020), dedicated (typically compensated) community management (hOurworld 2023, Buy Nothing Project 2023). Computation, as databases and verification algorithms, is often used to support micro level trust (i.e. a business and their customers), and in some cases replace trust with consensus algorithms based on the one-time expenditures of effort, time or resources (e.g. Bartoletti 2018). At larger scales, government intervention, interconnections with surrounding financial systems (e.g. Sardex), and tax statuses that require meticulous documentation to support revenue audits (e.g. TimeBanks) are all meso and macro level challenges. Thus the larger scale resists simple computational interventions;  additional schema design and record keeping within databases are required to avoid and support responses against intervening financial and legal institutions. In many senses community-based value exchanges challenge fiat money policy and therefore invite greater scrutiny.

---

[2] Valchokkryia et al. (2018) draw on de Certeau to define tactics as emergent approaches from negotiations between researchers and community organizations. This maps nicely to our metaphor of "root and water" process, in which both researchers and community organizations conduct explorations that "meet up" in the key locations for collaborative efforts (Eglash 2018).

*Democratic control of common-pool resources / Peer production*

In our work, we define democratic control of commons-pool resources as collaborative and participatory decision-making processes within a community that aims to prevent value extraction and maximize local value circulation. Following Ostrom (1990), to achieve successful cooperation, clear communication, trust, and reciprocity are required. Technology can play a crucial role in enabling this kind of cooperation. For instance, communication and peer production examples include various online forums, peer production platforms like GitHub, and others such as blockchain (Rozas 2021), the Mastodon social media app (Zia 2023; pg 10), and Decideum (Team 2023) for managing communication and reciprocity. However, while these and other technologies have been developed for commons-based deliberation, challenges remain. These include power imbalances, implicit and explicit biases, and violations of community guidelines ([Bauwens et al 2019a](#)). Therefore, it is crucial to establish clear guidelines, procedures, and mechanisms for conflict resolution and continually evaluate and adapt them to ensure that deliberative commons cooperation can be achieved sustainably and equitably. To do this there is a need to explore the prefigurative computing design's potential to enable bottom-up trade-offs between generativity and alienation.

*C. Summary*

The above examples suggest that there is some continuity between pre-computing efforts to develop community-based economies, and the new applications and innovations that technology has made available. The interplay between continuities and new affordances are helpful cautions against claims of a "techno-fix" (Huesemann and Huesemann 2011): the case studies show that it is as much about social innovation as technical. They also stress the need for attention to multiple scales and value flows. For example, putting a single worker-owned shop on Facebook might help maintain unalienated labor value for an individual business, but that simultaneously reinforces corporate platform dominance at the larger scale, taking up a niche that could have been occupied by a community-owned platform, with its greater potential for returns of unalienated social value.

The challenges are not restricted to the domain of private corporations. For example, Tandon et al. (2022) describe the state regulatory practices that created barriers to their computational support for a city-based taxi driver organization; regulations created by a sophisticated state lobby effort from Uber and Lyft. Such examples illustrate how the limited geographic scale implied by "community-based" can be an advantage: an economic system that is responsive to regional distinctions may be able to take a more democratic, environmentally attuned approach. But that comes with the caveat that other phenomena--assistance programs, multi-regional infrastructure, universal human rights, global warming and so on--require institutions at the state, federal and global level.

In addition to business sector challenges and public policy barriers, there can also be constraints that are more culturally embedded. The importance of prefigurative approaches in addressing these kinds of barriers can be illustrated by sociologist Sally Hacker (Hacker 1988) in her critique of sexism at Mondragon. If the founders are all male, attempting to add in feminist considerations decades later is less likely to be effective, compared to integrating it at the start. If we think of extractive formations like sexism, corporate domination, and so on as "basins of attraction"--patterns that can draw more resources into their grasp over time--then prefiguration acts as a kind of counterforce, helping community members develop their own self-sustaining formations that support cognitive and socio-environmental transitions away from those damaging basins of attraction (der Leeuew and Folke 2021; Eglash et al. 2014).

**Table 3 summarizes our background section by associating existing computational mechanisms with their fundamental solidarity strategies. Following that we articulate how our perspective on computing for community economies provides a novel synthesis using our guiding 3 principles outlined above.**

| Solidarity Sector | Facilitates … | Before computing | Subsequent computational form | Pros and cons |
|---|---|---|---|---|
| Mutual Credit Systems | Credit exchange (Gerber, 2015, Sartori, 2020, Scott, 1972, Schraven 2001) | Face-to-face interaction; paper based ledgers (eg papyrus to carbon copies); Patronage | Online platforms with: digital balance tracking, messaging, reputation systems; Sardex, | Opportunists or connection to opportunistic systems can tax and collapse the system |

| | | | Patreon | |
|---|---|---|---|---|
| Alternative Currencies[†] | Exchange of goods; bartering alternative ([Hileman](#), 2013, Ruddick et al., 2015, [Bristol Pound](#), 2023) | Physical tokens (beads, shells, scrip), commodity notes, scrip | Digital dollars, blockchain, digital platforms for face-to-face exchange (over Facebook, Craig's List, etc); Bristol Pound network, Etherum, Bitcoin | Often seen as illegal competition/use against national currencies; User trust may require currency redeemable to fiat; value alienation, external assurances |
| Time Banks | Labor exchange ([Dillahunt](#) et al., 2022, [Dittmer](#) 2013, [Cahn](#), 2000, [Valek](#), 2019) | Face-to-Face interaction; paper based ledgers; paid bank management; | Online platforms with: digital balance tracking, messaging, labor advertising, person-to-institution forms; [Time and Talents software](#) | Typically granted specific tax status that is delicate; Requires sustained attention towards long term vs short term relationships, peer to peer trust critical |
| Local Exchange Systems | Labor-Knowledge exchange, including (P2P, B2B) ([Privat](#) et al., 2020, [Williams](#), 1996) | Directory of wants, needs, and knowledge; paper based credit ledgers | Online platforms with: digital tracking, offer/need advertising; [Community Exchange System](#) | Typically organized as non-profit; Consolidation of online platforms (CES) |
| Peer Production | Combined knowledge/ labor exchange (Benkler and Nissenbaum, 2006a; [Bauwens](#) et al 2018; [Ramos](#) et al 2021; [Kostakis and Tsiouris 2024](#)) | N/A | Online platforms with: peer work coordination, aggregation, communication, offer/need advertising; Linux, Github, RepRap, OpenStreetMap | An incomplete form of production, taken up by surrounding economic systems (Kioupkiolis 2022); Community guidelines often implicitly and explicitly violated (eg sexism, racism in OS movements) |
| Sharing Economy | Worker-owned sharing economy, Mutual resource sharing ([Kimmerer](#) 2022, [Durkheim](#) (LibreText 2018), | Explicit within many indigenous practices, face-to-face interaction, ritual, social expectation, mechanistic solidarity practices | Online platforms with: offer/need advertising, work coordination, centralized routing algorithms; [Buy Nothing](#) groups | In digital form, often incomplete and taken up by and/or subsumed by surrounding contexts, such as platform capitalism; Can perpetuate cultural and spatial injustice |
| | Pasquale 2016, Tham 2022, Dillahunt et al. 2017) | | | |
| Worker Collectives | Self-organized labor (Rothschild-Whit 1979, Rothschild 2000; Trebor Scholz 2012) | Face-to-face interaction, written rule books, contracts, face-to-face negotiation, organization of ownership | Online platforms with: audio-video communication, voting systems, system of record, deliberation systems; drutopia.org | Hierarchical management colludes or becomes unresponsive;<br><br>Often targeted using reprisals |
| Socio-political support | Strong regional production of unalienated value (Vlachokyriakos et al 2018, Crivellaro 2019; Team 2023; Papadimitropoulos 2020) | Barcelona and the Basque region; Mondragon; Emilia-Romagna area of Italy. | Various online platforms | Typically requires historical support from governing bodies and other forms of solidarity (Zabin and Ringer 1997). |

Table 3: Summary of solidarity mechanisms and ways subsequent computation has been layered into them to create current community based economies.

† Credit vs Currency: a key difference between currency and credit is ownership. Currency is in-hand cash and credit is cash, virtual or actual, owned by someone else and under their terms (e.g. interest rates) IMB Bank (2022)

III. Methodology

We propose a collection of methodologies for empirical investigations of community-based economies (CBE), while providing participatory action research (McIntyre 2008) related prompts for grappling with multiscale effects — a resource for academic and non-academic researchers to broadly envision and foster technological CBE supports in ways they may not have considered them. Essentially, we carried out a qualitative, action-based, multi-case study approach. These cases were used to unpack thick connections between participants and evolving communities in context of ongoing CBE development. Three reflective action research prompts served as indicators, checks and balances: A) How can computing supports for CBEs and unalienated value circulation resist tendencies towards surrounding infrastructure; B) What applied principles and practices enable effective nesting of computational supports; C) How could impacts and implications differ across micro, meso, and macro scales of interaction? These prompts, along with our methodologies, guided our field experiments, ensuring an ongoing critical understanding of relationships between CBEs and technological supports. Table 4 indicates connections between these methodologies and reflective prompts, between approaches and broadly envisioned technological design and effects. These relationships are by no means absolute but suggest starting approaches to broadly envision and study computing for CBEs.

| Approaches | | Prompting for Broadly Envisioned Technological Design and Effects | | |
|---|---|---|---|---|
| Method | Forms | A) How can computational support for CBEs and unalienated value circulation resist tendencies towards surrounding infrastructure | B) What applied principles and practices enable effective nesting of computational supports | C) How could impacts and implications different across micro, meso, and macro scales of interaction? |
| Co-design | Discussing needs across business processes, proof of concept prototypes | X | | X |
| Qualitative Observation | Participant, Bystander, Interlocutor | X | | |

| Simulations | Visual virtual placement | | X | |
|---|---|---|---|---|
| Workshop | In-person convening, virtual convening | X | X | |
| Interview | Semi-structure, unstructured, colloquial; one time, ongoing | | X | X |
| Physical Fabrication | Rapid manufacturing technology, digital manufacturing | X | X | |
| Iterative Design | Field level user feedback and testing; reconceptualization of physical or virtual maker technologies | | X | |
| Speculative Design | Futuring sessions | X | X | X |
| Computer and Microprocessor Programming | Agile, Large Language Model assistants ("Co-pilots") | X | X | |
| Financial Assessment | Enterprise partnership, comparative analysis, pricing tool development | X | | X |
| Materials and Waste Assessment | Process analysis, comparative analysis | X | X | X |

Table 4: Methodological approaches and reflective prompts for envisioning computational support in CBEs

**A. Purposive Population Sample**

We collaborated with worker-owned enterprises in low-income communities, specifically those of an artisanal nature. The initial sample comprised 20 business owners, all of whom are African American, with 60% being female, as well as Ghanaian participants. This purposive sampling approach was adopted to align with our prefigurative objectives and ensure representation of historically marginalized groups in the research process. Geographically, we worked in Detroit city, a largely African American city with nearly half (41%) living below the federal poverty line, at the time of the study. Artisanal enterprises were chosen for their relationship to unalienated cultural value, relevance to the local economy, and collaborative capacity for fostering CBEs.

From this sampling we collaborated with participants to launch a wide variety of cases. For clarity we outline each case along with its approach and specific forms, in Table 5. In *A Multiscale Computing Infrastructure for Community-Based Economies* we argue and discuss how field evidence, derived from these approaches, relate case-based computational supports to the fostering of unalienated value circulations across micro, meso, and macro scales.

| Case | Method | Specific Forms |
|------|--------|----------------|
| African Futurist Greenhouse | Co-design; Qualitative Observation; Workshop | Co-design workshop with Kumasi Hive makerspace in Ghana; Collaboration with African Bead Museum in Detroit on design, hiring, training; Observation of construction process and youth technology training |
| Cornrow braiding simulations | Simulations; Co-design; Iterative | Simulations of braiding designs; Interviews with braiding shop owners about technology needs |

| | Design; Computer Programming | |
|---|---|---|
| Digital fabrication technology integration | Physical Fabrication; Qualitative Observation; Interview; Computer and Microprocessor Programming; Iterative Design | Experiments with 3D printing, laser engraving, direct-to-garment printing; Observation of technology use in artisanal production; Interviews about benefits, challenges, adaptations |
| Textile artisan projection system | Co-design; Simulations; Speculative Design; Iterative Design; Materials and Waste Assessment | Prototyping of low-cost projection setup for direct digital-to-fabric transfer; User testing and feedback from textile artisans |
| BatikBot development | Physical Fabrication; Speculative Design; Computer and Microprocessor Programming | Iterative prototyping of digital batik tool; Collaboration with batik artisans on design requirements |
| Open source clothing design software | Computer Programming; Financial Assessment; Iterative Design; Workshop | Initial adoption of Freesewing (De Cock, 2018) software for parametric design; Interviews with textile artisans about custom AI modalities; Prototyping of AI-enhanced Freesewing functionalities |
| Upcycled e-waste jewelry | Workshop; Physical Fabrication; Materials and Waste Assessment; | Interviews with artisans about current e-waste usage; Prototyping of AI-based recommendations for material substitutions and reuse |
| Payment and accounting services | Financial Assessment; Interview; Qualitative Observation; Computer Programming | Interviews with worker-owned businesses about financial challenges; Initial participatory design of electronic forms for benefits processing; Prototyping of community-owned payment processing solutions |
| AI-powered circular economy sourcing | Co-design; Materials and Waste Assessment; Computer Programming | Interviews with artisans about current material sourcing practices; Participatory design of AI app for waste stream and sustainable material discovery; Prototyping of dataset |

| | | integration and recommendation algorithms |
|---|---|---|
| Decidim platform governance | Speculative Design; Interview | Interviews with worker-owned businesses about governance needs; Comparative analysis of existing e-participation platforms |

Table 5: Overview of engagements, methods, and specific forms of community collaboration

Each case involved multi-month, and in some cases, multi-year, intensive collaborations, with Detroiters ranging from as many as 20 folks to a single person; as of this writing, some cases are still ongoing. Here we start at the grass roots, utilizing the prefiguration principle, to work with those already engaged in solidarity practices or are implicitly sympathetic to generative justice principles, without being familiar with the theory. These methodological choices set the groundwork for action research style community participation, iterative refinement based on field and in-situ feedback, providing grounded insights towards the opportunities of computing for CBEs for practice drawing from qualitative analyses.

**B. Connecting Multiscale Computing to Prefiguration, Generative Justice, and Solidarity**

Our purposive methods for computing mediate between a scale by facilitating information and resource flows sensitive to unalienated value, leading to new ways of social coordination and decision-making among communities that utilize these technologies. This sensitive coordination then shifts resource and process allocations across scales, thereby supporting via prefiguration the bottom-up, egalitarian organization of value generation and sharing in CBEs. At the micro scale, methods such as cornrow braiding simulations and digital fabrication foster new designs, exploration and production methods, increasing involvement in unalienated labor and prefiguring alternative modes of value co-creation. At the meso scale, tools for collaboration, resource sharing, and knowledge exchange strengthen community bonds and collective resilience among artisanal enterprises. Cases like the *African Futurist Greenhouse*, discussed in the following section, exemplify how technology can enhance solidarity practices within and between community groups, promoting generative justice through equitable exchanges of knowledge and meaningful forms of construction. These micro and meso-scale changes contribute to macro-scale shifts by challenging existing power structures and promoting systemic change, as seen in efforts to modify waste and money flows. Thus, with these methods, multiscale computing creates a synergistic effect across all levels, fostering a more equitable and sustainable economic ecosystem.

**IV. A Multiscale Computing Infrastructure for Community-Based Economies**

Below we describe what multiscale computing for community based economies might look like, by applying our methodology, we extend prior work as described in the background section to address the processes at micro, meso, and macro levels of analysis, and their cross-scale interactions. As a means of investigating this process empirically, our current NSF grant has allowed us to begin some experiments, which we integrate into the description below.

The three scales are not merely categories for descriptive convenience; they also describe nested functionalities that build on each other. It is common to see technical descriptions using such nesting: hardware to firmware to software in CS; biomolecules to cells to organs in physiology; soil microbes to photosynthesizers to food chains in ecology; and so on. Less frequent are understandings of social and economic justice as nested functionalities.

Previously we noted that historic cases of top-down communism tend to produce authoritarian state dominance, and historic cases of bottom-up free markets tend towards corporate dominance. The most common solution for this dilemma in the CS literature is participatory design, and its grounding in the phrase "nothing about us, without us". But Pelle Ehn (2014), a founder in participatory technology design, notes a consistent failure to provide large scale visions: instead the tendency is "many small capricious situated futures in the making" (p. 17). Responding to this challenge, Lodato and DiSalvo (2018) recommend larger scale technology design processes in terms of "institutioning" that resists neoliberal economic pressures. But the ultimate in institutioning could be said to be found in orthodox Marxist communism, and as we have noted that is vulnerable to the flaws of authoritarian collectivism (just as the free market is vulnerable to competitive individualism). It is for this reason that we approach the development of community-based economies as nested functionalities, in which the computational support at each scale--micro, meso, macro--is guided by our three principles: prefigurative development, generative justice, and solidarity economy.

First is the prefigurative approach, which is based on means-ends convergence. One vulnerability of participatory design is that the needs expressed in the short term may be preconditioned by dominant forces: for example facilitating use of a corporate-owned platform is both helping the user in the short run but strengthening the dominance of the corporation in the long run. Conversely, design fiction, speculative design and similar frameworks focus on futures, but that embeds a challenge to action in the present (Wong et al. 2020). Prefigurative design can help to bring these two approaches together through its focus on diverse participants, sustainable practices, democratic procedures, and other processes by which the means of development and the end goals of development are aligned.

Second is attention to value flows, such that the economy is minimizing extraction, and maximizing returns to the labor, social, and ecological sources of value generation (i.e. generative justice). And third, integrated across interacting domains, and across every scale, the free agency of individuals is balanced with opportunities for mutual aid. This "solidarity clause" is in one sense a restatement of generative justice, but it is often helpful to unpack the concept of "returns of unalienated social value" to illuminate its components of democratic process and cross-domain interactions with health, housing, education, environment, etc. (Appadurai 2001, Barber 2003).

Not every principle has equal weight at every scale. At the micro scale, in which we focused largely on small worker-owned business, we investigate how computing can support prefigurative diversity and unalienated labor value flows, with particular attention to digital fabrication, computational agroecology, and other "primary production". At the meso scale, where these businesses interact with each other and their consumers, we have fewer empirical results, but we describe some of our near-future plans for worker-owned cloud applications to facilitate solidarity principles. Commons-based peer production by sharing AI, pattern generation and other design apps, worker-owned delivery, and rerouting supply chains to develop a localized ecosystem that minimizes extraction are some of the features at this scale. At the macro scale we offer some visions for how community-wide processes such as governance platforms, and health, housing, education and environmental sustainability could arise from a community-based economy.

*A. Computing for micro-scale processes in community-based economies*

As noted above, we begin with a prefigurative process. In our case the "initial state" was provided by a focus on worker-owned enterprises in low-income communities in Detroit that are "artisanal" in nature. This offers four contributions to the prefigurative process:

1. As mentioned, Detroit is about 80% African American, with 41% of the residents living below the federal poverty line, as of this writing. By starting with this particular low-income community, we better enable a prefigurative process that aims for a future in which the needs of underserved demographics are prioritized.

2. Small "non-employee businesses" on average have more ownership by underrepresented groups in both race and gender (US Small Business Administration 2019). Thus the approach could be generalized to any location: starting with worker-owned enterprise will, on average, increase the odds for a prefigurative demographic. In our case the initial set of 20 enterprise owners are all African American, and 60% female.

3. Starting with worker-owned enterprises means that if financial empowerment is successful, we are prefiguring a future in which the majority of enterprises are worker-owned.

4. Despite what is often lower pay, artisanal enterprises are more likely to be sites of unalienated labor. Artisanal styles of work can be done in almost any domain, as long as they are doing what they love, with their own creative flair, styles of work, and other control over their own means of production (Garvin et al 2023, Luckman, 2015; Ocejo, 2017; Sennett, 2008; Solomon and Mathias, 2020). If we want a future of work in which people love what they do, a prefigurative approach recommends such a starting set. In our particular case, the occupations of the starting set of participants include agricultural education, arts education, barber, bespoke clothing, cosmetologist, craft artist, entrepreneurship education; food preparation and catering, food delivery, household furnishing fabrication, jewelry and adornment fabrication, landscaping, solar installation, and urban farming.

One need not have a singular focus for the prefigurative process. For example, feminist maker movements have made gender much more central than we have attempted here (c.f. (Okerlund, Wilson, and Latulipe 2021); likewise for ecology centered movements (Klemichen et al. 2022). Keeping an emphasis on democratic process in prefigurative development is one way to prevent any one focus from overwhelming other concerns.

General challenges at the microscale include the following:

1. Artisanal authenticity: pleasure derived from artisanal labor, and the attraction of the products for consumers, can be threatened by computational supports.
2. Physical economies of scale: many automation technologies are designed for larger scale operations. As Winner (1984) points out for the case of family farms wiped out by the automated tomato harvester, particular technology designs can over-determine political economy effects.
3. Material and human resource limitations: even when the above two barriers are resolved, additional limitations include typical development challenges: electrical, water, healthy air venting for new machinery, covering costs, training, etc.

We are especially interested in what the computing research community can bring to bear on these challenges. The innovation deficit for low-income communities, ensuring that public research dollars are routed toward benefiting wealthy corporations is self-reinforcing: funding agencies such as the NSF, in our experience, are less likely to fund experiments in emerging technologies (AI, robotics, etc) for grass roots operations due to the 3 challenges above, making them into self-fulfilling prophecies. For this reason we have highlighted in the above sections some of the strategies that have broken this cycle, such as Newcastle University's Open Lab, which emphasizes the solidarity economy, and FabCities, which emphasizes makerspace production. Our own approach highlights generative justice and its circulation of unalienated value. Within that framework, returns of cultural capital, Indigenous knowledge and other forms of "heritage algorithms" (Bennett 2016) has been an especially important feature for computational support, and that has continued in the Detroit project. Here the integration of cultural capital -- discovery and re-discovery of ancestral practices -- facilitate micro-scale approaches between.

For example, urban farming represents some unique opportunities for technological innovation; combining digital sensing with automated regulation of growing parameters in water, nutrients, heat and so on in particular. As a means of integrating cultural capital, we developed the African Futurist Greenhouse (Eglash 2020). In a workshop at Kumasi Hive makerspace in Ghana, local members developed simulations of traditional african architecture. In collaboration with the African bead museum in Detroit, we co-developed a plan for local design input, hiring, training and product distribution. Local community members, some with disabilities that limited employment, were hired for on-the-job construction training. Youth interns in our workshops developed automated digital controls. The thermal advantages allowed growing indigo for dyes, African seeds sold as beadwork, and other artisanal production contributions as well as healthy local food for consumption. Photos of the entire process can be viewed at https://generativejustice.org/afg. More general cases for urban farming will be discussed in the meso-scale section. The return of unalienated value, and in particular the integration of cultural capital, was critical to both acceptance and community-based innovation: had we merely said "greenhouse technology that optimizes productivity" it might have been rejected at the outset, or specialized in products that maximized profit rather than support for an artisanal economy.

Heritage algorithms and related support for the integration of cultural capital also facilitated some of the micro-scale interventions in digital fabrication, primarily as means of addressing challenge #1, "remixing" expectations of computational supports by demonstrating support for overlooked cultural practices. These include 3D printing, laser engraving, and direct to garment printing. These often combine with other practices as hybrid forms, such as 3D printing to create a mold that is filled with another material (either metal such as jewelry, biodegradable material to enhance sustainability, or edible substances for consumption). Like the greenhouse example above, in some cases we began with simulations: for example cornrow braiding designs that had been used for computing lessons, and later used to create unique 3D printed mannequin heads for braiding shops (Lachney et al. 2021). The simulation connections to STEM offered additional returns to the community in the form of culture-based education, as well as enhancing revenue streams for participants who included education as part of their service sales.

Other products were unrelated to cultural specifics. Looking across all cases, the digital fabrication technology was distributed along the familiar use-adaptation-innovation spectrum (Bolosha et al., 2022; Eglash, 2004; Kirton, 1976). The main advantages for digital fabrication at the microscale at this point are provisional (small numbers of examples, or not fully implemented), but the potentials can be categorized as follows:

1. Increased repertoire of products and services. Creative advantages include the ability to utilize materials previously unavailable, explore patterns in simulations before physical rendering, and to supplement creativity with AI assistance.

2. Improvements in labor process, such as speed, cost, reduced tedium, and sustainability of feedstocks.

3. Systemic effects such as open source sharing and other new opportunities to connect to other artisans (discussed in meso-scale section), feeling better connected to contemporary trends in technology and markets, and so on.

As noted in challenge #2 above, while digital fabrication is well supported in the context of financial wealth--industrial factory, university machine shop, etc.--adaptation is often needed to address resource barriers in low-income circumstances. Some are due to unconventional processes (eg using recycled wood for laser engraving). Others are adaptation to insufficient infrastructure, such as venting in a basement. Some are cost-saving: for example we mounted a projector on a ceiling to allow a textile artisan to do away with paper tracing and go directly from digital screen to cloth (avoiding higher-cost commercial systems that do the same). Others are better classified as innovation: for example a BatikBot creating a digital version of hand crafting. Our initial experiments using AI at this micro-scale of labor process also fall along this spectrum. For example, we are currently combining FreeSewing, an open source software project for parametric clothing design, with unique AI applications for the tasks prioritized by our textile artisans.

In summary: of the three challenges, #3, resource limitations to expand technology access (e.g. fabrication infrastructure) might be the most difficult to address. The emphasis on makerspaces we see in some community technology development visions (Okerlund et al. 2021) can be helpful in terms of providing opportunities to learn the technologies, but it is also a limiting model compared to permanent installation and ownership of the means of production by workers. Focusing exclusively on narrowly defined technical goals for CS research, such as "improving usability", can likewise divert attention to the kinds of innovation and adaptation that goes beyond profit-making and contributes to community-generating processes. The development process itself, not just the technology, needs to be co-designed with the community members if we are to contribute to transformative pathways that create egalitarian and sustainable community-based economies (Dillahunt et al., 2023). And even with co-design, the assumptions of extractive economic relations are deeply embedded in everything from urban infrastructure to financial exchange systems. There is an obduracy (Hommels 2008; 2020) and technological momentum (Mondschein et al., 2020) that resists changing socio-technical relations, making deeper reform so costly that we often settle for shallow workarounds or band-aids.

*B. Computing for meso-scale processes in community-based economies*

At the meso-scale, digital platforms have increasingly dominated economic exchanges, diminishing the possibilities for local economic independence (Johnson and Eglash 2021). More generally, when online services exist for localized production, such as Etsy for crafts, DoorDash for food, or Uber for transportation, the platforms employ economies of scale, transaction costs, restrictive APIs and other competitive market mechanisms that monopolize access and limit local development (Burrell and Fourcade 2021; Hoppner 2022; Peck and Phillips 2020). In particular, the dominance afforded by these platforms amplify wealth inequality, and with that increased stratification by race and other social impairments (Rahman 2019). The work on platform cooperatives (Zhu and Marjanovic, 2021) has made strong progress in demonstrating how platforms owned and supported by the workers themselves is a promising alternative. The framework of community-based economies may be useful in extending that concept to encompass linkages between multiple platform cooperatives (Vidal, 2022). But doing so will require addressing specific counters to the domination effects listed above (Kioupkiolis, 2022).

In our initial experiments towards the mesoscale, we have collaborated with some of Detroit's worker-owned small businesses to explore three AI and server-side applications, with an eye towards these counter-hegemonic needs. Our first technology addresses the challenge of high costs and fees associated with platform economic dominance between businesses and their customers. This is partly because showing immediate payoff for participants is important for their continued engagement: if we claim there will be benefits, but only after four years, it will diminish interest. Here our research explores analogous but community sensitive functionalities offered by dominant platforms for free or at lower cost on their own platform, to allow workers to retain more value. One possibility is payment and accounting services. Developing electronic forms to streamline the processing of paper documents for benefits like SNAP, EBT, and WIC is one area. Another area is credit card processing fees, which regularly extract percentages of profit while demanding complex technological maintenance and delays in payment clearing. But, perhaps due to developing digital payment infrastructures, banking in the hands of those that are not bankers can offer payment processing without percentages and faster payment clearing, through certified money service business as fintech companies providing banking as a service.

Ongoing research also investigates additional strategies to better suit the needs of worker-owned collectives. Some might be relatively simple:  improve these payment interfaces to have greater ease of use and utility for the lay people at the helm of these enterprises. Another possibility is physical distribution networks through a routing application for cultural, social, and ecological sustainability; again it will be critical to examine policy and institutional contexts (e.g. Tandon et al. 2022). For example there are opportunities to include driver social preferences, along with cultural and ecological multiplex routing algorithm integrations using the open source routing machine codebase (Giraud 2022) while working with Detroit's urban farmers. Some of these functionalities are prohibitive: for example we will not attempt to launch a community-owned server farm for custom machine learning processing, as that is simply not a "low hanging fruit" for our team (although it could become a recommendation result from our research). But our preliminary interviews revealed many services that are both costly to vendors, and relatively easy to implement in a bespoke manner.

Circular and cultural economy building at the mesoscale has to combine the above financial concerns--how to keep worker-owned business afloat--with other forms of value that prevent alienation in ecological and social domains. Through applications that promote lateral creative thinking between humans and AI, we seek to facilitate cultural and ecological synergies, such as the upcycling and recycling of materials for socially beneficial purposes. One area of our exploration involves an app that utilizes AI-powered datasets (extant data analyzed by natural language processing to facilitate client side inferencing), to deliver information on sourcing from waste stream materials, ecologically sustainable biomaterials, or other circular economy resources.

Our test case is the utilization of copper wire from e-waste, which our Detroit groups currently carry out by manual selection (U-M Stamps School of Art & Design, 2023). An AI-based service might for example expand that to include wire coat hangers from dry cleaners, and recommendations for replacement processes and products to handle the difference between copper and steel. That is to say, AI-based sourcing may go beyond geographic locating functions, and develop more creative collaborations as hybrids in which both the conceptual space of design possibilities and the physical space of waste stream resources are combined. By combining AI capabilities, specifically large language models, with human input and refinement, we aim to create a platform that fosters sustainable practices through the sharing of multiple knowledge forms. The intertwined concepts of material sourcing and sustainability life cycle analysis coupled with the creativity of lateral thinking can promote holistic approaches to resource utilization and waste reduction where it might have not existed.

Lastly, in discussing the meso scale, we also consider the concept of intransitive material networks, drawing inspiration from ecological models. Intransitive food "rings" in ecosystems, where species form interconnected relationships without a clear hierarchy, have been found to correlate with lower extinction rates and better biodiversity (Alcantara et al. 2017). Similarly, in local material supply chains, we can explore the idea of deliberate specialization and collaboration among worker-owned companies to reduce competition. By using AI to facilitate this collaboration and encourage businesses to focus on complementary materials and their lifecycle, we can create an meso level materials ecosystem that aligns with the intransitive model. By moving away from products and focusing on materials we attend to the important legal considerations regarding monopolies and price fixing (U.S. Code 1890; U.S. Office of Public Affairs 2015), and emphasize the need for a supportive sociopolitical infrastructure for community-based economies to flourish. In a sense, this is a materialized version of Piore and Sabel's "flexible manufacturing networks," discussed in 2.b.iii, but inspired by a lower layer of production than products and empowered by AI and human view of material flows to support economies. Other views of this concept, such as the flexibility gained when the intransitive ring of materials does not close, suggest that the functional diversity (Mason et al. 2005) supported may be the greatest benefit.

In conclusion, the meso scale presents unique challenges and opportunities for community economies between businesses and customers. Together by addressing cost burdens, promoting circular and cultural economy building, exploring networks, and leveraging open-source software and AI applications, businesses can establish and strengthen flows of unalienated value.

*C. Computing for macro-scale processes in community-based economies*
When circumscribed by our community economies platform, participatory design decisions at the macro scale arise from ongoing consensus making among groups of participants belonging to different cultures and subcultures, typically in a committee or esteemed council like setting. Three notable concerns include:

1. Platform governance: the labor and maintenance for the platform can be significant while requiring a synthesis of strongly held and divergent perspectives.

2. Deliberative consumption: in a community-based approach, what products are made and are purchased require a thoughtful approach (Wildenberg et al. 2013) between production processes, consumers and the environmental and social impacts.

3. Availability of and access to information: Search engine algorithms are increasingly subjects of critique, with evidence indicating their role in driving polarization, exclusion, and algorithmic social harms, and so a more participatory approach may be possible, with those made vulnerable by algorithmic unfairness having a voice in how they want to be "found."

Although there is a relatively large literature on democracy and design, as well as one on e-participation platforms (see review in Augusto and Baek 2021), the studies specifically on platform cooperative governance are smaller in number. But they all stress the importance of this feature (Schneider, 2020), as it must balance the long-term sustainability of business with the needs of worker self-determination: a tension at the core of the concept of a democratized economy. Governance software such as Decidim, designed with platform cooperatives in mind, may be most appropriate. Smith and Martín (2022) note that Decidim or its related variants are used by over 130 institutions in 33 countries, ranging from city governments to energy corporations. Aragón et al.

(2017) contrast Decidim with systems such as e-Petitions Gov UK, which allows comments to be categorized (for, against, neutral), but no threaded discussions, and Consul, which allows threaded discussions but no categorization of comments, Decidim combined the two. As a result they were able to classify threaded discussions by the degree of "cascades" of replies, and to show a correlation with negative comments, suggesting that new information was being supplied (i.e. counter-arguments), and thus richer deliberation process.

For deliberative consumption, the challenge lies in transforming the relationship between consumers and producers within the community-based economy paradigm. Much attention has been given to the "prosumption" concept of user-driven customization, but this is typically understood within a mass-production framework (von Hipple 2016; (Comor 2010)). A community-based approach, in contrast, would strive for "deliberative consumption" (Wildenberg et al. 2013) in which relationships between bespoke production methods, consumers and the environmental and social impacts are more thoughtfully composed, and better supported by technological facilities, within the platform. Another opportunity arises from "passive" methods of fundraising for platform sustainability. For example, because software developers often prefer API access to web-scraping, there is also an opportunity here to "passively" generate income through small fees using blockchain technology (low carbon versions only) or digital payments. Platform support for educational outreach from community-based enterprise can also cycle value forms, as technology uptake by local entrepreneurs can inspire STEM aspirations for local youth (Eglash et al. 2017; Lachney et al. 2021), and conversely payments for their involvement in schools can provide them with additional funding streams.

Finally, and crucially, data and technology access requires ongoing consideration, both proactive and reactive. Beyond information sharing and collaboration via forums, external information architectures, such as search engine indices and back linking practices, raise the question of how platform resources should be found by others and who those others are. A key paradigm here is generative searching. By adopting a participatory design process in the development of search engine algorithms we can better mitigate algorithmic unfairness and bias. And by involving vulnerable communities, such as African American artisanal entrepreneurs from low-income backgrounds, in the design process, their voices and needs can shape how they want to be "found" online. Other ongoing investigations have shown that through a mixed methods approach qualitative narratives can be translated into computationally accessible forms to ensure that search results prioritize themes and perspectives valued by the community, fostering greater relationality and addressing algorithmic bias. We label this approach as "content aware." By centering participatory search engine criteria and recognizing the importance of inclusivity and community input, the computing research community can contribute to more equitable and representative online experiences.

Reflecting on these challenges, the greatest limitation is the labor and maintenance required to sustain the platform itself. However, the recursive governance design process presents an opportunity to experiment with automated content moderation, reducing burdens, and enhancing efficiency. These key takeaways together with a governance council emphasizes the importance of ongoing participatory approaches and the use of recursive governance to drive positive social change and sustainable community economies at the macro scale.

V. An Operationalization Thereof

Operationalization is the process of converting abstract concepts into concrete methods and measurements. In the prior sections we have pointed to empirical field investigations to describe how multi-scale computing infrastructures have unfolded in ways that strive to construct community-based economies in a co-participatory manner. To support designers, both academic and non-academic, in their deliberative evolutionary efforts, we propose a guiding combination of methods that can serve as a starting point. While researchers or designers can begin with these principles, it is the participants who must carry them forward in their own application and work in multiscale ways. Networking and fostering workshops are excellent ways of scaffolding initial growth. Before presenting the operationalization, we want to establish well known sources for applying each method along with critiques thereof. For co-design, we refer to the work of (King et al. 2022; Zamenopoulos and Alexiou 2018), with a particular note on the critiques of the co-design workshops as participatory design (Huybrechts, Benesch, and Geib 2017). For speculative design, we refer to the work of (Bray et al. 2022; Elsden et al. 2017; Harrington and Dillahunt 2021; Harrington, Klassen, and Rankin 2022). For rapid development, agile or alternative strategies may be especially relevant (Abrahamsson et al. 2017; Cockburn 2006; Hron and Obwegeser 2022). In terms of measurement, assessing prefigurative progress in terms of means-ends convergence, determining whether iterations are leading in the intended direction, is one way visible to many.

To incorporate the solidarity principle, workshopping is a way to intersect and unite community members through focused discussions, assessments, and free-wheeling conversations. This approach has a long history and critiques, as described in (Charlotte Smith et al. 2020; Dourish et al. 2020; Harrington, Erete, and Piper 2019; Lazem et al. 2022; Udoewa 2022; Winschiers-Theophilus, Zaman, and Stanley 2019; Wyche 2021) . For generative justice, the paradigm that centers ownership of the means of production to achieve unalienated value circulation, introductions and extensions can be found in (Eglash 2016) and (Eglash et al. 2020), particularly among agentic justice (Bennett 2021). In our work, we introduced basic principles of unalienated value generation and circulation; and the critique of value extraction and alienation. We showcased them through various technological, bioreactive, and other demonstrations related to the participants current work. This prompted discussions among participants, translating the concepts of generative justice as they relate to their own experiences and aspirations. Together with the references, a designer can embark on deliberative evolution by selecting methods from Table 2 and Table 3 and periodically assessing progress through the measurement columns. Succinctly put, the methods in Table 2 can be viewed as what to do and Table 3 as where to foster it with participants.

**A. Table 2: Multiscale Computing Infrastructure Design: Methodological Operationalization**

| Principle | Method | Method | Method | Measurement |
|---|---|---|---|---|
| Prefiguration | Co-design | Design Fiction | Rapid and novel application development | Degree of convergence using the means-end principle, from mixed method assessment |
| Solidarity | Workshops | Industry Focus Groups<br><br>E.g. design groups, farmer groups | Cross-sector Groups<br><br>E.g. multiple stakeholders groups, across occupations | New and increased opportunities for mutual aid, from mixed method assessment |
| Generative Justice | Framing discussions that introduce Generative Justice paradigm | Identification and discovery of alienating and alienating value pathways | | Reduction in (a) ratio of  and (b) changes in alienated and unalienated value source/sinks, from mixed method, interpretivist assessment |

**B. Table 3: Multiscale Computing Infrastructure Design: Locating Methods among Scale Level**

| Scale Level | Method | Measurement |
|---|---|---|
| Micro | Participant sense making<br><br>Participant developing Artisanal Futures related viewpoints | Ongoing structured and unstructured interview<br><br>New and changed business relationships |
| Meso | Group sense making<br><br>Group developed Artisanal Futures related viewpoints | |
| Macro | Multi-group developed Artisanal Futures related viewpoints | |

Positive roles for the contemporary state and government in the development of community-based economies can be found in many of the examples we have discussed above. For instance, in Barcelona the city council worked with community-based economy advocates from FabCity and other groups to launch the Decidim participatory democracy platform, allowing more funding and policy support for localized and sustainable production (Aragón et al., 2017; Giovannini 2020; Wouter 2022). In 2019 the governor of California signed a bill allowing counties and municipalities to establish public banks, which will mean that the enormous profits normally created by private mortgages and loans will be going instead into public ownership; a virtuous cycle that can fund less profitable ventures in public housing and solar.   At the other extreme, Tandon et al. (2022) found that state policy was preventing the formation of localized transportation cooperatives, because of lobbying efforts by Uber and Lyft. In our own Detroit efforts to tie sustainability to low-income community empowerment, we found that the ability for photovoltaics to run power meters in reverse--to have power companies pay for local solar energy contributions--had been blocked by corporate lobbying. Creating state policy, laws and institutions that enhance generative justice, and removing those that block it, is fundamental to the model we propose. The goal is to create a balanced approach that recognizes the state's presence while working towards a more self-sufficient and resilient model of community-based economies that can thrive alongside, and eventually beyond, the state.

Conclusion
Transition strategies can be thought of as a spectrum between purely opportunistic--what might be called "guerilla infrastructuring" in which current systems are appropriated or co-opted (e.g. squatting, DIY, solar punk, and similar movements)--and purely visionary, in which a finalized design is preconceived and imposed in a grand gesture of social engineering, as was prominent in the 19th century social utopians like Charles Fourier and Robert Owen.  Building, for example, a worker-owned marketing platform to replace Etsy, is quite different from merely co-opting current infrastructure (setting up stores on corporate platforms), in which the community becomes structurally dependent, not only on the infrastructure, but also ongoing auxiliary harms the infrastructure is linked to. This is because infrastructure is constituted by persistent co-creation acts that themselves entail extraction. Even a counter-hegemonic move like "creative reuse", such as applying waste to improve ecological sustainability, or donating SMS credits to improve financial sustainability, can unintentionally support the extraction that is causing the excess in the first place (planned obsolescence for physical waste, promotional programs for SMS, and so on). On the other hand, insisting on purity is often impractical: it is challenging to compete with corporate solutions where profitability is linked to economies of scale and extractive advantages, and the time and resources to build large scale infrastructure may be substantial and cause lengthy delays.

We refer to our mid-spectrum approach as "deliberative evolution" because it combines democratic consideration of long-term principles with short term experiments. The short term experiments can be analogous to (and in some cases the literal embodiment of) the agile software development process (Dybå and Dingsøyr 2008), in which frequent trials of incremental changes and contextualized small scale prototypes (Deininger et al. 2019) are used to gather formative feedback. We take the term "deliberative" from "deliberative democracy" (Gutmann and Thompson, 2004)), which emphasizes thoughtful public discussion and grassroots decision-making. The 3 principles we have outlined here--prefiguration, generative justice, and solidarity--are in this sense produced through deliberative conversation with agile experiments. It will be up to the community participants themselves (including non-human agents in nature, which can speak loudly in fires, floods and crop failures, but have positive and softer tones as well) to determine the tradeoffs.

Kelty (2005) refers to open source as a "recursive public", because unlike a public library or civic plaza, the citizens can be both makers and democratically governed inhabitants of the code's utilities and structures. But we only "inhabit" code as one part of our existence, and dominant forces have become experts at externalizing costs to the other parts, creating the illusion of liberation online and off-loading destruction and exploitation to physical existence. Community-based economies are aimed at closing that gap, at providing "full stack decolonization" (Eglash et al. 2023) that creates a recursive public across all domains.

Our empirical case studies at micro, meso, and macro scales illustrate how deliberative evolution works at and between scales, facilitating information and resource flows sensitive to unalienated value. Micro-scale examples like cornrow braiding simulations demonstrate how technology can enhance traditional skills and foster unalienated labor. At the meso level, projects such as the African Futurist Greenhouse show how community-based enterprises leverage technology to strengthen local resilience and knowledge sharing. Macro-scale initiatives to modify waste flows exemplify systemic challenges to existing economic structures. These diverse cases highlight how communities can leverage multiscale computing to promote intricate unalienated connections using deliberative evolution. While the number of case studies may seem limited, their depth and diversity offer rich insights into the synergistic effects of multiscale computing, supporting our larger claim that multiscale computing can and has fostered egalitarian community-based economies.

It is thus in the deliberative evolution process that we can seek solutions to competition from corporations, and their advantages from leveraging economies of scale and extraction itself. As we shrink the sources of extraction, and expand unalienated value returns, the process can reveal new challenges -- too difficult given our time constraints and resources; others do-able but fail to meet participant approval, cost, security, or other criteria -- as well as new solutions. It is only in the act of carrying out the research, guided by these principles, that multi-scale computing support for community based economies can reveal these deliberative evolution trajectories.

Acknowledgement

This material is based upon work supported by the National Science Foundation (NSF) Grant No. 2128756. All opinions stated or implied in this document are those of the authors and not their respective institutions or the National Science Foundation.